\documentclass{elsart}
\usepackage{natbib,graphicx}

\begin{document}

%%%%%%%%%%%%%%%%%%%%%%%%%%%%%%%
\def\sun{\hbox{$\odot$}}
\def\la{\mathrel{\mathchoice {\vcenter{\offinterlineskip\halign{\hfil
$\displaystyle##$\hfil\cr<\cr\sim\cr}}}
{\vcenter{\offinterlineskip\halign{\hfil$\textstyle##$\hfil\cr
<\cr\sim\cr}}}
{\vcenter{\offinterlineskip\halign{\hfil$\scriptstyle##$\hfil\cr
<\cr\sim\cr}}}
{\vcenter{\offinterlineskip\halign{\hfil$\scriptscriptstyle##$\hfil\cr
<\cr\sim\cr}}}}}
\def\ga{\mathrel{\mathchoice {\vcenter{\offinterlineskip\halign{\hfil
$\displaystyle##$\hfil\cr>\cr\sim\cr}}}
{\vcenter{\offinterlineskip\halign{\hfil$\textstyle##$\hfil\cr

>\cr\sim\cr}}}

{\vcenter{\offinterlineskip\halign{\hfil$\scriptstyle##$\hfil\cr

>\cr\sim\cr}}}

{\vcenter{\offinterlineskip\halign{\hfil$\scriptscriptstyle##$\hfil\cr

>\cr\sim\cr}}}}}

\def\degr{\hbox{$^\circ$}}
\def\arcmin{\hbox{$^\prime$}}
\def\arcsec{\hbox{$^{\prime\prime}$}}
\def\utw{\smash{\rlap{\lower5pt\hbox{$\sim$}}}}
\def\udtw{\smash{\rlap{\lower6pt\hbox{$\approx$}}}}
\def\fd{\hbox{$.\!\!^{\rm d}$}}
\def\fh{\hbox{$.\!\!^{\rm h}$}}
\def\fm{\hbox{$.\!\!^{\rm m}$}}
\def\fs{\hbox{$.\!\!^{\rm s}$}}
\def\fdg{\hbox{$.\!\!^\circ$}}
\def\farcm{\hbox{$.\mkern-4mu^\prime$}}
\def\farcs{\hbox{$.\!\!^{\prime\prime}$}}
\def\fp{\hbox{$.\!\!^{\scriptscriptstyle\rm p}$}}
%%%%%%%%%%%%%%%%%%%%%%%%%%%%%%%%%%%%%%%%%%%%%%%%%%%%%%%%%%%%%%%%%%%%%%

\newcommand{\D}{\displaystyle} %normal formulas
\newcommand{\T}{\textstyle} %for large font
\newcommand{\SC}{\scriptstyle} %footnote
\newcommand{\SSC}{\scriptscriptstyle} %footnote to footnote

\newcommand{\ttbs}{\char'134}

\newcommand{\AmS}{{\protect\the\textfont2

  A\kern-.1667em\lower.5ex\hbox{M}\kern-.125emS}}

\def\AJ{{\it Astron. J.} }
\def\ARAA{{\it Annual Rev. of Astron. \& Astrophys.} }
\def\ApJ{{\it Astrophys. J.} }
\def\ApJL{{\it Astrophys. J. Letters} }
\def\ApJS{{\it Astrophys. J. Suppl.} }
\def\ApP{{\it Astropart. Phys.} }
\def\AA{{\it Astron. \& Astroph.} }
\def\AAR{{\it Astron. \& Astroph. Rev.} }
\def\AAL{{\it Astron. \& Astroph. Letters} }
\def\JGR{{\it Journ. of Geophys. Res.}}
\def\JHEP{{\it Journal of High Energy Physics} }
\def\JPhG{{\it Journ. of Physics} { G} }
\def\PhFl{{\it Phys. of Fluids} }
\def\PR{{\it Phys. Rev.} }
\def\PRD{{\it Phys. Rev.} { D} }
\def\PRL{{\it Phys. Rev. Letters} }
\def\Nature{{\it Nature} }
\def\MNRAS{{\it Month. Not. Roy. Astr. Soc.} }
\def\ZA{{\it Zeitschr. f{\"u}r Astrophys.} }
\def\ZFN{{\it Zeitschr. f{\"u}r Naturforsch.} }
\def\etal{{\it et al.}}

\hyphenation{mono-chro-matic  sour-ces  Wein-berg
chang-es Strah-lung dis-tri-bu-tion com-po-si-tion elec-tro-mag-ne-tic
ex-tra-galactic ap-prox-i-ma-tion nu-cle-o-syn-the-sis re-spec-tive-ly
su-per-nova su-per-novae su-per-nova-shocks con-vec-tive down-wards
es-ti-ma-ted frag-ments grav-i-ta-tion-al-ly el-e-ments me-di-um 
ob-ser-va-tions tur-bul-ence sec-ond-ary in-ter-action
in-ter-stellar spall-ation ar-gu-ment de-pen-dence sig-nif-i-cant-ly
in-flu-enc-ed par-ti-cle sim-plic-i-ty nu-cle-ar smash-es iso-topes 
in-ject-ed in-di-vid-u-al nor-mal-iza-tion lon-ger con-stant
sta-tion-ary sta-tion-ar-i-ty spec-trum pro-por-tion-al cos-mic
re-turn ob-ser-va-tion-al es-ti-mate switch-over grav-i-ta-tion-al
super-galactic com-po-nent com-po-nents prob-a-bly cos-mo-log-ical-ly
Kron-berg Rech-nun-gen La-dungs-trenn-ung ins-be-son-dere
Mag-net-fel-der bro-deln-de}
%Formeln
\def\simle{\lower 2pt \hbox {$\buildrel < \over {\scriptstyle \sim }$}}
\def\simge{\lower 2pt \hbox {$\buildrel > \over {\scriptstyle \sim }$}}

% instructions for automatic equation numbering
%Benutzung:  $$ Formel\eqno\autnum$$

\journal{New Astronomy Reviews}

\begin{frontmatter}

\title{Cosmic rays, stellar evolution, and supernova physics}

\author[1,2]{Peter L. Biermann}
\address[1]{Max-Planck Institute for Radioastronomy, Bonn, Germany}
\address[2]{Department for Physics and Astronomy, University of 
Bonn, Germany}
\ead{plbiermann@mpifr-bonn.mpg.de}

\begin{abstract}
Most cosmic rays are thought to be accelerated by the shocks of supernova
explosions of very massive stars.  Here we review one quantitative
proposal, which predicted the spectral slopes, bend and cutoff about the
cosmic ray spectrum across the spectral bend or ``knee" near $3 \,
10^{15}$ eV in 1993.  Many of the specific predictions have now been
verified, and so it may be appropiate to investigate the consequences of
that proposal.  The successful fit to the cosmic ray data across the 
knee  suggests: 1) very massive stars, above about 20 to 25 solar
masses all converge to a common final state; 2) the supernova explosion
of these very massive stars is caused by a combination of rotation,
magnetic fields and the gravitational potential, just as proposed by G.
Bisnovatyi-Kogan in 1970, based on an earlier suggestion by N.S.
Kardashev in 1964; 3)  their stellar winds as well as the explosion
provide the main injection of magnetic fields into the interstellar
medium;  4)  the explosive energy of these supernovae is about $10^{52}$
erg; 5)  the explosion is extremely anisotropic, which may unify several
classes of observed supernova types; 6)  gamma ray bursts may be the
ultimate version of such an explosion in the case the collapse leads to a
black hole; 7)  it is plausible that the luminosity of the supernova
integrated over all aspect angles is also the same for all these massive
star explosions, leading to 8) a possibly very bright standard candle, if
we could just find a correction for the extreme asphericity.  This may
solve the problem of what the mechanism of supernova explosions is for
very massive stars, where most cosmic rays and magnetic fields come from,
and may also point the way to a unifying scheme for supernova explosions
and gamma ray bursts. 
\end{abstract}
\end{frontmatter}

\section{Stellar evolution and Supernova explosions}

Massive stars form, evolve and finally explode as supernovae.  The lower
mass limit for single stars that explode is estimated to be near 8 solar
masses.  Stars in the mass range 8 to about 15 solar masses evolve
without a strong stellar wind, and so explode directly into their
environment, in the now classical picture of a self-similar explosion
(e.g. D.P. Cox 1972).  More massive stars, between about 15 and about 20 -
25 solar masses explode as red supergiant stars (RSG) with a wind.  These
winds are moderately powerful, and the surface of the star slowly exposes
the deeper layers of Helium, so that the wind has a strong admixture of
Helium at the time of explosion.  From about 20 - 25 solar masses the
stars evolve to a blue supergiant stage, the Wolf Rayet stage, with very
powerful winds.  These winds eat back into the star exposing the deep
layers, and so show a mixture of Helium, Carbon and Oxygen.  These winds
are very strong, and produce a massive wind-shell, a shell of snow-plowed
interstellar material, and former wind material, which in turn is of
course former outer stellar material.  The driving of these winds already
had been a problem; it is generally accepted that radiation drives the
wind, and that wave excitation transfers the momentum; the coupling
constant is the wave speed.  The temperature of the wind and so its speed
of sound is strongly limited by cooling to 10,000 to 20,000 degrees K. 
Following earlier work by J. Cassinelli and St. Owocki we showed some
time ago, that magnetic fields could help this problem by showing that
the coupling between radiation and wind goes through the dominant wave
speed, which could be the Alfv{\'e}n speed; magnetic waves are not
influenced by cooling.  The optimum momentum transfer appears to occur
when the wind is slightly super-Alfv{\'e}nic, with an Alfv{\'e}nic Mach
number of a few, say, 3 (work with H. Seemann, 1997).  

Wolf Rayet stars may hold the clue to our understanding of high energy
cosmic ray particles, the origin of magnetic fields, the physics of
supernova explosions, the origin of gamma ray bursts, and may even give
us a bright standard candle for cosmology.

A much more extended recent version of some of these arguments is in two
reviews by P.L. Biermann \etal $\;$ (2003a, astro-ph/0302168; 2003b,
astro-ph/0302201), two reviews based on talks at the Palermo meeting Sep
2002; these reviews also contain many important references, which are
omitted here for lack of space.

\section{ The cosmic ray spectrum}

Cosmic ray particles, discovered in balloon flights just after 1910 by
V.F. Hess and W. Kohlh{\"o}rster, are now known to extend in particle
energy to a few times $10^{20}$ eV.  Around $3 \, 10^{18}$ eV there is a
kink in the spectrum and also a change in the chemical compositon, and
this is generally believed to be the transition between Galactic cosmic
rays and extragalactic cosmic rays:  However, a final proof for such an
interpretation is still waiting to be discovered.  Here we concentrate on
the cosmic ray particles below $3 \, 10^{18}$ eV, and ask what their
sources may be.

So let us first summarize their properties:

The spectrum of cosmic rays is approximately $E^{-2.7}$ until the
knee, which is a bend downwards at around $3 \, 10^{15}$ eV.  The
spectrum beyond the knee is approximately $E^{-3.1}$.  There is a
slight downward dip from $3 \, 10^{17}$ eV, and a transition near $3 \,
10^{18}$ eV.  At $3 \, 10^{18}$ eV there is switch from a quite steep
local slope to a much flatter slope, as well as an apparent change in
chemical composition from medium to heavy nuclei to light nuclei
(Hydrogen and Helium).  A general review including (almost) all data up to
1997 is in B. Wiebel-Sooth \& P.L. Biermann (1999).

The key features for Galactic cosmic rays are the high particle energies,
and the spectral bend at the knee; we also need to understand the second
``knee" at $3 \, 10^{17}$ eV, where the spectrum dips to merge into the
extragalactic cosmic rays near $3 \, 10^{18}$ eV.  Berezinsky \etal $\;$
(2003) have recently discussed a very similar explanation of this dip.

Supernova explosions into the interstellar medium do not give such a
kink, and also are not usually accepted to give a maximum particle
energy of $3 \, 10^{18}$ eV (P.~O. Lagage \& C.~J.  Cesarsky 1983).  The
kink at around $3 \, 10^{15}$ eV has had a variety of proposals to
explain it:  The main proposal has been the notion that it reflects a
change in propagation:  This entails a steepening of the energy
dependence of the propagation by about 0.3 to 0.4 at a specific
energy/charge ratio $E/Z$.  Such a  specific energy/charge ratio $E/Z$
corresponds to a precise length scale in the interstellar medium, about
0.4 parsec.  At that length scale the properties of the irregularities in
the interstellar medium should change rather drastically.  There is no
such evidence.  Also, if that were the case, the anisotropies near
$10^{18}$ eV should be large, and from the AGASA data at that energy we
know that the anisotropy is small.  All available evidence suggests that
as regards cosmic ray transport the interstellar medium can be described
by a Kolmogorov law from very small scales much below 0.4 pc all the way
up to much larger scales, up to a few hundred pc.  This is still a
fraction of the scale of the hot gas, magnetic field and cosmic ray disk,
which has a full width of about 4 kpc, as suggested by ROSAT data
(Snowden \etal $\;$ 1997).  Therefore we conclude that the explanation of
the kink has to be sought in the source.

In a series of papers (first qualitatively, H.J. V{\"o}lk \& P.L. Biermann
1988, and then quantitatively in several papers starting with P.L.
Biermann 1993) we have suggested that the key to understand the
acceleration of higher energy particles was the recognition that many
massive stars explode into their own wind.  In a magnetic wind, such as
the Solar wind (L.F. Biermann 1951), the magnetic field has a basic
topology already suggested by E.N. Parker (1958), with $B_r \sim 1/r2$,
$B_{\theta} = 0$, and $B_{\phi} \sim \sin \theta /r$, an Archimedian
spiral.  Therefore the dominant magnetic field runs as $1/r$ with radial
distance $r$, and so the Larmor radius of any particle with some given
energy/charge ratio $E/Z$ scales linearly with $r$.  Any shock racing
through the wind introduces also a scale from the snow-plow effect, of
$r/4$.  Near the pole the acceleration is in a parallel magnetic field
configuration (i.e. shock normal parallel to the locally prevailing
magnetic field), which is the situation discussed by P.O. Lagage \& C.J.
Cesarsky (1983): near the equator the acceleration is in the near
perpendicular configuration as discussed by R. Jokipii (1987); the
perpendicular configuration introduces also drift acceleration, while
both configurations also need to include adiabatic lossses, and the
density history of injection.  The key is the introduction of the notion
of the ``smallest dominant scale" to describe the large scale turbulence
in the unstable shock region, both in real space as in velocity space
(P.L. Biermann 1993); the character of the turbulence in the shock region
is the critical argument here, as recognized by many.  The final result
is the predicted spectrum of $E^{-8/3 - 0.02 \pm 0.02}$ below the knee,
$E^{-3.07  - 0.07 \pm 0.07}$ above the knee, for wind-SNe; from matching
the two acceleration regimes the knee energy, as well as the ankle
energy, are $E_{knee} = Z e B(r) r (3/4 U_{sh}/c)2$, and $E_{ankle} = Z e
B(r) r $.  The gradual cutoff with $Z$ gives the second knee, and then the
steepening between $3 \, 10^{17}$ eV and $3 \, 10^{18}$ eV.  Both
quantities, $E_{knee}$ and $E_{ankle}$, are constant with $r$ in a
magnetic wind, considering most of $4 \pi$.  Putting in numbers suggested
in 1993 (see the review at the Calgary 23rd ICRC meeting, P.L. Biermann
1994) that the knee and ankle energies are $0.6 \, Z \, 10^{15}$ eV, and
$Z \, 10^{17}$ eV, with a fair uncertainty due to a specific choice of
numbers for stellar properties, that are hard to know.

Recently, with the advent of the KASKADE data we have been able to test
the predictions, and determined in a direct fit the two spectra to
$E^{-2.67}$, and $E^{-3.14}$, and the two energies to $E_{knee} = 1.7 \,
Z \, 10^{15}$ eV, and $E_{ankle} = 2.2 \, Z \, 10^{17}$ eV.  A full scale
Monte-Carlo CORSIKA fit to all vertical and slanted shower data available
from KASKADE gives very similar numbers, but then with error bars:  The 
two spectra are $E^{-2.65 \pm 0.03}$ and $E^{-3.25 \pm 0.04}$, and the
knee energy comes out to $2.1 \pm 0.14 \, Z \, 10^{15}$ eV, while the
ankle energy cannot be easily determined in this way, using a Monte-Carlo
approach; all details are described in P.L. Biermann \etal $\;$ (2003b). 
Therefore the predictions were confirmed quantitatively.

This implies that all sources must be very similar, because otherwise a
source based model could not describe an overlap of many injection
events.  Since the asymptotic Parker magnetic field topology in our
context implies that the rotation of the star is highly differential so
as to allow a highly tangential field already at the surface, this
finding from the fit also has consequences for the rotation of the star,
since angular momentum loss is minimized.

\section{Origin of magnetic fields}

These stars and their winds are magnetic, and so inject magnetic fields
into the interstellar medium.  We noted above that the Alfv{\'e}nic
Machnumber is most likely of order 3.  This entails that the termination
shock of the wind injects magnetic fields into the environment that have
already near 10 \% of equipartition values.  Considering the origin of the
magnetic field in the interstellar medium, and its topology, this may be
all that is necessary in terms of strength - however, this does not
produce any order in the field, and that requirement leads to very
different considerations.  One might ask what magnetic fields massive
stars produce in analogy to the Sun:  The convective interior will
produce a magnetic field from the seed field (derived from the fact that
surfaces of constant pressure and constant density do not generally
coincide in a rotating system, and so drive an electric current, L.F.
Biermann 1950, L.F. Biermann \& A. Schl{\"u}ter 1951) and the dynamo
mechanism (M. Steenbeck \& F. Krause, and E.N. Parker).  This magnetic
field is quite strong, and can readily be transported to the surface by
meridional circulations, also known as Sweet-Eddington circulations. 
Therefore we conclude that magnetic field may be relevant  in the
evolution of massive stars, as well as to explain the interstellar
magnetic fields.  As galactic winds and radio galaxies transport these
galactic fields into intergalactic space, such as clusters of galaxies,
and sheets and filaments of the cosmic web of the galaxy distribution, it
might just be that Wolf Rayet stars hold one of two keys to explain all
magnetic fields in the universe - the other key is the explanation of the
symmetry and order.

\section{Implications for the supernova mechanism}

If the fit says that all stars give the same knee and ankle energy, then
this immediately suggests that explosion energy, magnetic field, and
rotation are deeply connected, and are in fact always nearly the same
number for each exploding star, above the zero age main sequence star mass
of about 20 - 25 solar masses.

This strongly supports the concept proposed by G. Bisnovatyi-Kogan in
1970, that the explosion of these massive stars is caused by a
combination of rotation, magnetic fields and potential energy, based on
an an earlier suggestion by N.S. Kardashev in 1964.  The idea is as
follows, and reminds one immediately of many much more recent arguments
about gamma ray bursts (especially the work by C. Wheeler and S.
Woosley):

The core of the star rotates, and so when it collapses due to a lack of
supporting pressure, it contracts into a small flat disk in rotational
balance.  This is at a few $10^{7}$ cm.  Since at that size the magnetic
field would exceed the Landau level, this transfer of energy implies many
rotation periods of energy injection during the approach to this size. 
The magnetic field is all sheared and wound up, and so transports angular
momentum and the potential energy of the small disk to the outside,
exploding the star.  The energy of the rotating small disk constitutes
the visible energy of the supernova, and is readily estimated to be
$10^{51}$ to $10^{52}$ erg.  When the small disk then finally collapses
into a neutron star or into a black hole, then that last energy is
radiated away as neutrinos, as observed.

This mechanism obviously is extremely anisotropic, and it can safely be
predicted that supernovae following this concept appear very different if
viewed from different angles.  It is quite plausible that various
supernova types might be unified in such a scheme, very much in analogy
to the unified scheme long used in active galactic nuclei, where aspect
angle is a major parameter.  It can be expected that along the symmetry
axis explosion speeds might be very much faster than along the equatorial
belt, and so may explain various observations of very fast knots.  It is
not clear at this point just what the precise connection to gamma ray
bursts could be; if this is also the general mechanism for gamma ray
bursts, in addition to the main mechanism for massive star explosions or
supernovae, then gamma ray bursts could be very much more frequent than
commonly argued.

Now considering the transfer of energy and angular momentum from the small
collapsed disk by magnetic fields, it is clear that this is fully
symmetric along the rotation axis between ``up" and ``down", giving a
saddle point instability. This entails, that the slightest asymmetry
between the two sides will give rise to a powerful sling shot effect
exactly along the rotational and symmetry axis, leading to a high speed
pulsar, which flies just along the direction defined by its own rotational
axis.  This can be tested observationally, and would be a second
powerful confirmation of the mechanism of G. Bisnovatyi-Kogan (see
Bisnovatyi-Kogan \& Moiseenko 1992).

Also, there is another corollary:  Using the abundances in the
pre-supernova winds (taken from N. Langer) we can determine the
connection between cosmic ray flux for some element like Carbon, and
connect the appropiate supernova rate with the energy contained in cosmic
rays for each single event.  This leads to an estimate of order $10^{51}$
erg for each event in cosmic ray particles alone, and then using an
inefficiency of 0.1 to an explosion energy of $10^{52}$ erg, already quite
close to what has been called a Hypernova by B. Paczynski.

\section{Gamma ray bursts}

There is growing evidence that gamma ray bursts and supernovae might be
connected (see, e.g., Uemara \etal $\;$ 2003, Price \etal $\;$ 2003,
Hjorth \etal $\;$ 2003, and many other comments and articles on GRB
030329).  There are two simple possibilities:  First, for extremely
massive stars the supernova explosion might lead to a central black hole,
and maybe just those stars also produce a gamma ray burst.  In this case
the frequency of gamma ray bursts might be much higher than one per
million years in a normal galaxy.  Second, it is also conceivable that a
binary system is required, with the main star exploding triggering a
gamma ray burst in the small neutron star companion leading to a collapse
to a black hole.  In this case the mass transfer towards the neutron star
companion could dramatically increase just prior to the explosion, and so
trigger a catastrophic collapse very close to the supernova explosion. 
In the first case the gamma ray burst would happen at exactly the same
time as the supernova, with very high velocities along the line of sight,
and in the second case the gamma ray burst would have to precede the
actual supernova explosion, because otherwise the supernova explosion
itself would dirty the gamma ray burst with baryonic matter beyond
recognition.  Maybe Nature realizes both options.  The literature shows
that there are many more possibilities.

A magnetic mechanism provides an interpretation of the extreme degree of
gamma ray polarization found recently (Coburn \& Boggs 2003, GRB 021206).

If we could find a way to observe the traces of the last few gamma ray
bursts in our Galaxy, then the time scale could be strongly restricted. 
In recent work with R. Engel, G. Medina-Tanco and G. Pugliese we have
proposed that the AGASA excess of events at $10^{18}$ eV in the Galactic
Center region is such a trace.

\section{Consequences for cosmology}

To emphasize the conclusion from the cosmic ray data fit:  The explosion
energy has to be very nearly the same for each such star, as also the
rotation, magnetic field and mass - all measured just prior to the
explosion.  The stellar evolution of these stars leads to a common final
state just prior to the explosion.

This suggests that the light curve of the supernova may also be the same,
if integrated over all aspect angles.  Such an integration might be
possible using infrared emission, echoing and polarization.  This remains
to be worked out and tested.  On the other hand, if this correction for
anisotropy could be done, then we would have a standard candle for use in
cosmology which has the promise to be much brighter than the commonly used
supernova type Ia.

\section{Conclusion}

Wolf Rayet stars are one key to understand the magnetic fields in the
universe, the enrichment in heavy elements, and much of the observed 
cosmic rays; cosmic rays in turn suggest that the magneto-rotational
mechanism by G. Bisnovatyi-Kogan does explain the explosion of these
stars, possibly providing a key to understand gamma ray bursts.  The
cosmic ray data imply that these stars converge to a common final state
in their evolution, just prior to the explosion.  Finally, if we could
work out the corrections for anisotropy, these supernovae could be very
bright new standard candles in cosmology.

\section{Acknowledgements}

PLB would like to acknowledge working on these questions most recently
with S. Casanova, R. Engel, N. Langer, H.S. Lee, G. Medina-Tanco, S.
Moiseenko, A. \mbox{\O}deg\mbox{\aa}rd, G. Pugliese, E.-S. Seo, S.
Ter-Antonyan, and A. Vasile. He would also like to acknowledge very
lively and useful discussions on these topics with I. Axford, Z.
Berezhiani, V. Berezinsky, G. Meynet, G. Sigl and T. Stanev.  These
results were first introduced in this breadth at lectures in Paris spring
2002 and then at the Palermo meeting September 2002 (see below, on
astro-ph), and PLB would like to express his appreciation to N. Sanchez
and her colleagues for their continuing hospitality.  Work with PLB is
mainly being supported through the AUGER theory and membership grant 05
CU1ERA/3 through DESY/BMBF (Germany); further support for the work with
PLB comes from the DFG, DAAD, Humboldt Foundation and the
Naumann-Foundation (all Germany), grant 2000/06695-0 from FAPESP (Brasil)
through G. Medina-Tanco, KOSEF (Korea) through H. Kang and D. Ryu, ARC
(Australia) through R.J. Protheroe, a NATO-grant with S. Moiseenko
(Russia), and European INTAS/ Erasmus/ Sokrates/ Phare grants.  Finally,
PLB is very grateful to the organizers of the meeting at Seeon May 2003,
specifically R. Diehl, for the invitation to present these results.  The
author also wishes to thank S. Moiseenko for a critical reading of the
article.

\end{document}